\documentstyle[12pt,psfig]{article}                    
\begin{document}
\title{Effective Functional Form of Regge Trajectories}
\author{\\ M.M. Brisudov\'{a},\thanks{E-mail: BRISUDA@T5.LANL.GOV} \
L. Burakovsky\thanks{E-mail: BURAKOV@T5.LANL.GOV} \ and \ 
T. Goldman\thanks{E-mail: GOLDMAN@T5.LANL.GOV} \
\\  \\  Theoretical Division, MS B283 \\  Los Alamos National Laboratory \\ 
Los Alamos, NM 87545, USA }
\maketitle
\begin{abstract}
We present theoretical arguments and strong phenomenological evidence that 
hadronic Regge trajectories are essentially nonlinear and can be well 
approximated, for phenomenological purposes, by a specific square-root form.  
\end{abstract}

\bigskip

\section{Introduction}
It is well known that the hadrons composed of light $(n\equiv (u,d),s)$ quarks
populate approximately
linear Regge trajectories; i.e., the orbital momentum $\ell $ of the 
state is proportional to its mass: $\ell =\alpha'M^2(\ell )+\alpha (0),$ 
where the slope $\alpha'$ depends weakly on the flavor content of the 
states lying on the corresponding trajectory. Therefore, knowledge of Regge 
slopes and intercepts is very useful for spectral purposes. Since knowledge of
Regge trajectories in the scattering region $(t<0)$ is also useful for many 
nonspectral purposes, e.g., in the recombination and fragmentation models,
Regge trajectories become a valuable description of hadron dynamics,
perhaps generally more important than the mass of any particular state. 

In the Veneziano model for scattering amplitudes \cite{Ven} there are 
infinitely many excitations populating linear Regge trajectories. The same 
picture of linear trajectories arises from a linear confining potential 
\cite{KS} and the string model of hadrons \cite{BN2}.

However, the realistic Regge trajectories extracted from data are {\it 
nonlinear.} Indeed, the straight line which crosses the $\rho $ and $\rho _3$ 
squared masses  corresponds to an intercept $\alpha _\rho (0)=0.48,$ whereas 
the physical intercept is located at 0.55, as discussed below. The nucleon 
Regge trajectory as extracted from the $\pi ^{+}p$ backward scattering data is 
\cite{Lyu}
\begin{eqnarray}
\alpha _N(t)=-0.4+0.9\;\!t+\frac{1}{2}\;\!0.25\;\!t^2,
\end{eqnarray}
and contains positive curvature. Recent UA8 analysis of the inclusive 
differential cross sections for the single-diffractive reactions $p\bar{p}
\rightarrow pX,$ $p\bar{p}\rightarrow X\bar{p}$ at $\sqrt{s}=630$ GeV reveals 
a similar curvature of the Pomeron trajectory \cite{UA8}:
\begin{eqnarray}
\alpha _P(t)=1.10+0.25\;\!t+\frac{1}{2}\;\!(0.16\pm 0.02)\;\!t^2.
\end{eqnarray}
An essentially nonlinear $a_2$ trajectory was extracted in ref. \cite{Bol2} 
for the process $\pi ^{-}p\rightarrow \eta n.$

In addition to being disfavored by these experiments, linear trajectories also 
lead to problems in theory. Linear trajectories violate Cerulus-Martin fixed 
angle scattering bound \cite{GM}. (The so called square-root trajectory, used 
extensively in this paper for spectroscopic purposes, saturates this bound 
\cite{Schmidt}.) Further, they violate  the Froissart bound, in the following 
sense: The consequences of the strong duality between the saturation of the 
$S$-matrix by i) narrow resonances \cite{FS} and ii) Regge asymptotic behavior
\cite{JT} in the $S$-matrix formulation of statistical mechanics by Dashen, Ma
and Bernstein \cite{DMB} was studied in \cite{JTS}. There it was shown that in 
order that the total cross section satisfy the Froissart bound \cite{Fro}, 
$\sigma _{tot}(s)\leq A\ln ^2s,$ $A={\rm const,}$ the density of resonances in
the elastic amplitude should decrease with energy, typically as $\sim 1/E$
\cite{JTS}. (A decreasing density of resonances is a typical feature of dual
amplitudes with Mandelstam analyticity (DAMA) \cite{Jenk}.) A linear 
trajectory violates the requirement of a decreasing density of resonances (for
a linear trajectory, the density of resonances grows as $\sim E^3$ 
\cite{cube}), and therefore, violates the Froissart bound in this framework. 

The idea of nonlinear Regge trajectories is not new. In the late-60's, by 
introducing Regge cuts, through the 
eikonal method, a number of authors have shown that the effective trajectory 
for large momentum transfer $(-t)$ goes like $\sqrt{-t}$ \cite{60s}. 
Subsequent comprehensive analysis by Vasavada, combining both Regge poles and 
cuts, arrived at the effective square-root trajectory in the entire complex 
angular-momentum plane \cite{Vas}. The $\alpha (t)\sim \sqrt{-t}$ trajectory 
was also found by Gribov {\it et al.} \cite{Gribov} in a Pomeron-reggeon field
theory. Other nonlinear forms have been  studied  also in the literature.
 
Once the nonlinearity of Regge trajectories is an established fact, the 
determination of its actual form becomes an important issue. 
In this paper we address this issue. We start in Section 2 with a model study 
of a heavy quark-antiquark system in a potential fitted to the results of an 
unquenched lattice QCD \cite{BBG}, and show that its (parent) Regge trajectory 
is equally well fitted by both limiting cases of nonlinear forms allowed by 
dual amplitudes with 
Mandelstam analyticity (DAMA) \cite{DAMA} (and thus, one can argue that all 
these nonlinear forms would fit the trajectory as well. The two limiting cases 
are the so called ``square-root'' trajectory and the ``logarithmic'' 
trajectory.)  To gain some understanding of this peculiar observation, we turn 
to analytic model of a massless string with a variable tension (Section 3).
(We show that this model is a relativistic generalization of a nonrelativistic 
rod with arbitrary potential, in the same sense as the usual Nambu-Goto 
string model is a generalization of a nonrelativistic linear potential model.) 
In this model we are able to recover underlying potential (or, in other 
words, the intergal of the variable string tension) from a known Regge 
trajectory, and we present the two potentials corresponding to the square-root 
and logarithmic trajectories, respectively, and compared them with the 
unquenched lattice QCD potential \cite{lattice1} used in the heavy quark model.
The two models (Section 2 vs. Section 3) are very different, but since the 
nonlinearity of Regge trajectories arises due to the flux tube breaking 
\cite{BBG}, one can expect the same {\it qualitative} behaviour in both heavy 
and light quark systems.

Sections 2 and 3 show that even though the two extreme nonlinear forms can 
both fit the bound states spectra, one of them, the sqaure-root form, goes 
beyond the position of the few lowest lying poles. This is why we choose the 
square-root form to fit and predict real-world spectra in Section 4. It should
be noted that any nonlinear form bracketed by the square-root and logarithmic
ones can be expected to give comparably good results for the lowest lying 
states, and a true test would be higher excited states (likely yet not 
observed).

With a few additional assumptions that are commented on as they are introduced,
we obtain an excellent agreement with data. This means that the Regge 
trajectories are indeed nonlinear, and well approximated by the square-root 
form. We summarize and comment on some of the results at the end of Section 4.
The last section contains our overall summary and conclusions.

\section{Heavy quarkonia model}
In a previous paper \cite{BBG} we considered a Hamiltonian model for heavy 
quarks which attempted to include the effect of light pair creation, or, in 
other words, the breaking of the flux tube stretched between the two heavy 
sources that inevitably occurs when the distance between the sources is 
sufficiently large. We have shown that consequently, Regge trajectories for 
the heavy quarkonia become nonlinear and their real parts terminate. In this 
paper we use the model to gain insight about the specifics of the (nonlinear) 
form of hadronic Regge trajectories.

A meson consisting of two heavy quarks is well described in leading order by a
nonrelativistic, spin-independent Hamiltonian, viz.
\begin{eqnarray}
H = -{1\over{2m}}{\boldmath {\nabla}^2 }+ V(r),
\end{eqnarray}
where $m=M/2$ is the reduced mass, with $M=5.2$ GeV, and  $V(r)$ denotes
a potential. We use $M=5.2$ GeV, and a potential fit to results of unquenched 
lattice QCD calculation for infinitely heavy sources. We have argued that our 
simple two body calculation produces results qualitatively similar to coupled 
channel method, because it dynamically takes into account coupling to open 
channels, and that the two approaches can be expected to differ significantly 
only for a few states near the threshold \cite{BBG}. 

The screened static potential fitted to results of lattice calculations
is~\cite{lattice1}:
\begin{eqnarray}
V(r) = \left( -{\alpha \over{r}} +\sigma r\right) {1 -e^{-\mu r} \over{\mu \, 
r}},
\end{eqnarray}
where $\mu ^{-1}=(0.9 \pm 0.2)$ fm $=(4.56 \pm 1.01)$ GeV$^{-1}$, 
$\sqrt{\sigma}=400$ MeV and $\alpha =0.21 \pm 0.01$. 
We obtain bottomonium-like  spectra by diagonalizing the Hamiltonian (3)
with the potential (4). The number of bound states in the model is finite, 
and it decreases with increased screening $\mu$. The number of bound states is
the largest for the parent trajectory, and decreases by one unit for each 
consecutive daughter. Another important feature of the Regge trajectories in 
this model is that they acquire curvature due to screening of the linear 
potential. Since we are interested in how the 
screening affects the form of the Regge trajectories (which would be linear in 
the absence of screening), for maximum reliability of the numerical fits it is 
useful to consider the case of a Regge trajectory with the largest number of  
bound states. Therefore, in what follows we concentrate on the parent 
trajectory for $\mu ^{-1}=0.9$~fm $=0.18$~GeV$^{-1}$. The same conclusions as 
those presented below hold for all $\mu$ in the range indicated by the fit to 
the lattice data \cite{lattice1}.

A meson trajectory $\alpha _{j\bar{i}}(t),$ can 
be parametrized on the whole physical sheet in the following form: 
\begin{eqnarray}
\alpha _{j\bar{i}}(t)=\alpha _{j\bar{i}}(0)+\gamma \Big[ 
T_{j\bar{i}}^\nu -(T_{j\bar{i}}-t)^\nu \Big] ,\;\;\;  0 \leq \nu \leq 
{1\over{2}}.
\end{eqnarray}
(up to a power of logarithm), assuming that  $\alpha_{j\bar{i}} (t)$ is 
analytic function having a physical cut from $t_0$ to 
$\infty ,$ it is polynomially bounded on the whole physical sheet,
and there exists a finite limit of the trajectory phase
as $|t|\rightarrow \infty $ \cite{Tru}. Here\footnote{Note
that the choice of the signs in Eq. (5) is fixed by the requirements
that $\alpha (t)$ must be real as $t\rightarrow -\infty$, and have
positive slope in the small-$t$ region.}  $\gamma $ is the
universal slope which nonlinear trajectories must have in asymptopia 
\cite{CJ}, $$\alpha (t)\sim -\gamma (-t)^\nu ,\;|t|\rightarrow \infty ;$$
both $\gamma$ and the exponent $\nu$ are flavor independent. $\alpha 
_{j\bar{i}}(0)\leq 1,$ in accord with the Froissart bound.
Note that for 
$\vert t \vert \ll
T,$ Eq. (5) reduces to the (quasi)linear form
\begin{eqnarray}
\alpha _{j\bar{i}}(t)=\alpha _{j\bar{i}}(0)+\nu \gamma T_{j\bar{i}}^{\nu -1}
t=\alpha _{j\bar{i}}(0)+\alpha'_{j\bar{i}}(0)\;\!t.
\end{eqnarray} 
The subscripts $i,j$ indicate which of the parameters depend on the quark 
content of the meson. In this section  we drop the subscripts, since our model 
is applied here to  a bottomonium-like system only.
\begin{figure}
\vskip0.25in
\centerline{\psfig{file=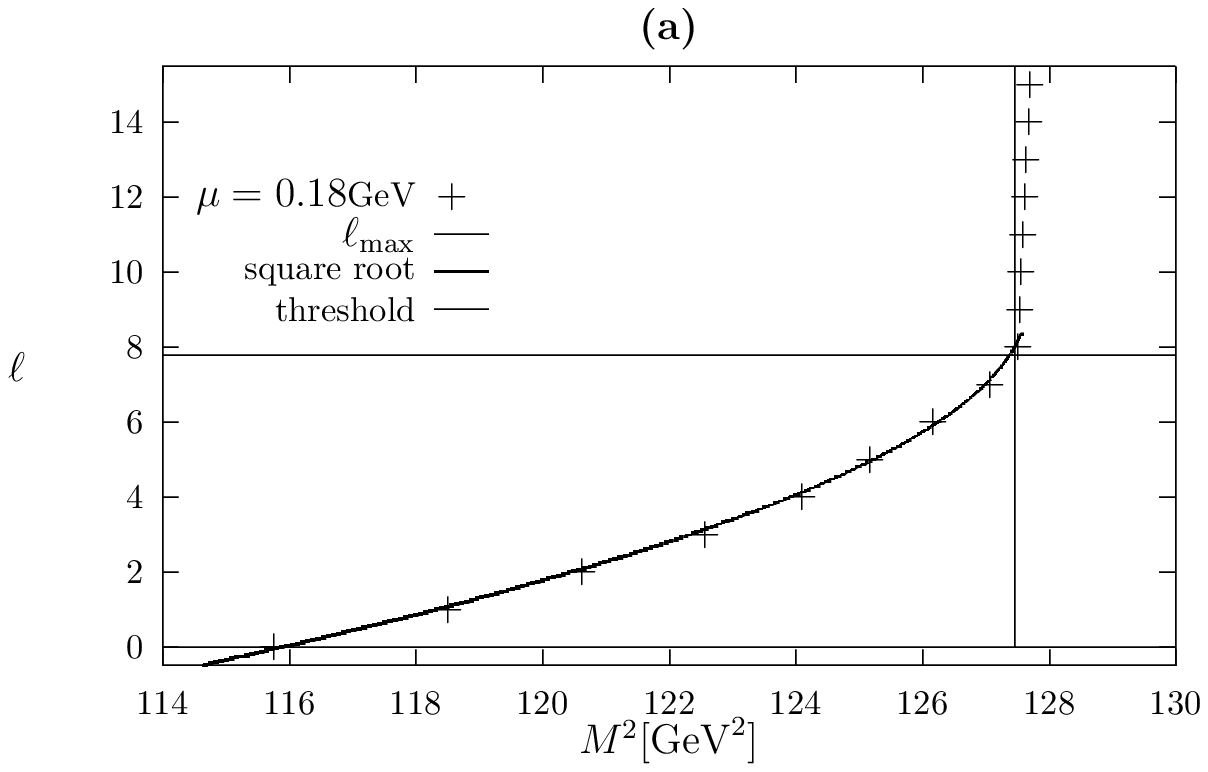,width=6in}}
\vskip0.25in
\centerline{\psfig{file=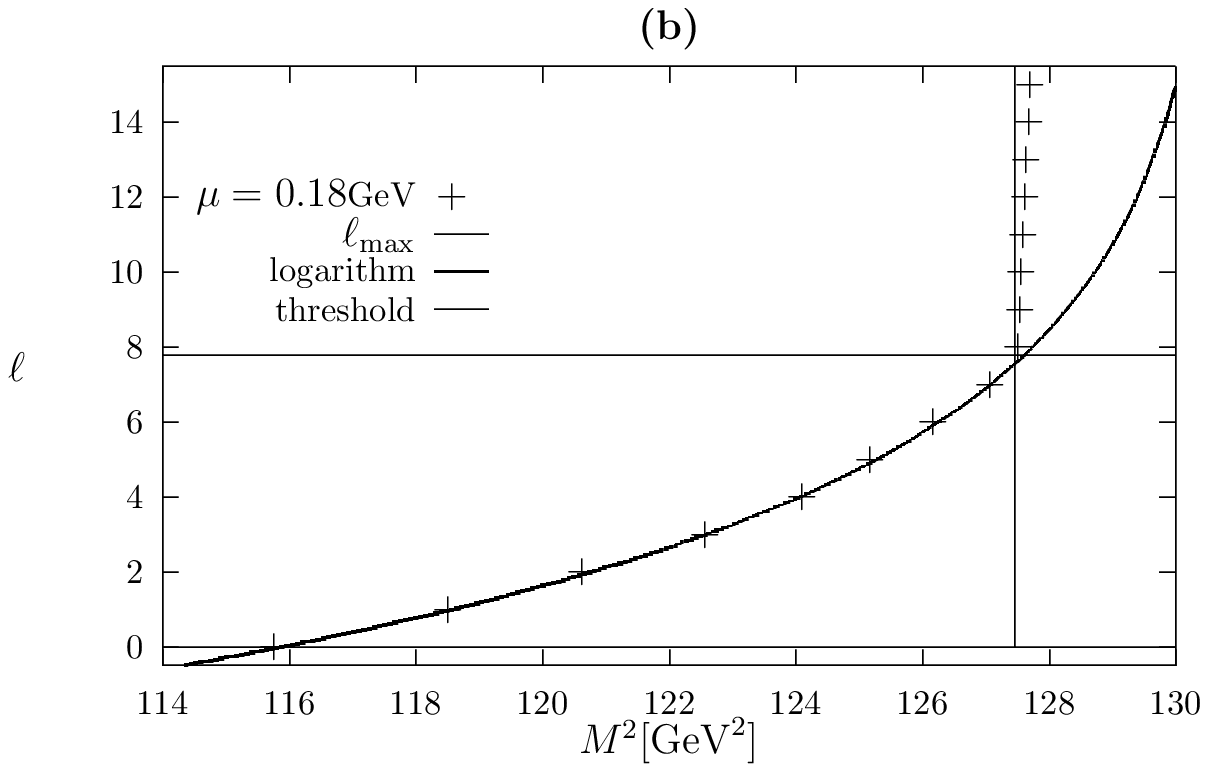,width=6in}}
\caption[]{The parent trajectory for the bottomonium in our model compared to 
{\bf (a)} our best fit of the ``square-root'' form, 
{\bf (b)} our best fit of the ``logarithmic'' form. The solid vertical line 
corresponds to the threshold of the quenched lattice potential, i.e. 
$(2M+\sigma / \mu)^2$, the solid horisontal line shows the maximum $\ell$ for 
the data set.}
\end{figure}
We consider the value of $\nu$ restricted to lie between $0$ and $1/2$ 
(\cite{LASZLO}, for details see Appendix A). $\nu=0$ should 
be understood as a limit $\nu \rightarrow 0$, $\gamma \nu $ fixed. In this 
limit, the difference of fractional powers reduces to a logarithm, viz.,
\begin{eqnarray}
\alpha (t)=\alpha (0)-(\gamma \nu ) \log \left( 1-\frac{t}{T}\right) ,\;\;\;
\gamma \nu = {\rm const.}
\end{eqnarray}
Unlike a trajectory with $\nu \neq 0$, the real part of the ``logarithmic'' 
trajectory does not freeze-out when $t$ reaches $T$. The real part continues 
to grow, the only change  for $t>T$ is that the trajectory acquires a constant 
imaginary part.

The upper bound on $\nu$ gives the so-called ``square-root'' trajectory, viz.
\begin{eqnarray}
\alpha (t)=\alpha (0)+\gamma \Big[ \sqrt{T} -\sqrt{T-t} \Big] .
\end{eqnarray}
When $t$ reaches $T$, the real part of the ``square-root'' trajectory stops 
growing, and there are no states with a higher angular momentum than
$ \ell_{max} =\Big[ \alpha (T)\Big]$. The parameter T is therefore the 
trajectory 
termination point.

Fig. 1(a) shows the parent Regge trajectory for our model together with our 
best fit of the ``square-root'' form:
\begin{eqnarray}
\alpha_{\sqrt{\ } }(0)& = &-21.12\pm 7.74, \nonumber\\ 
\gamma_{\sqrt{\ }}& = &2.68\pm 0.97 \ [{\rm GeV}^{-1}], \nonumber\\
T_{\sqrt{\ }}& = &127.66 \pm 1.98 \ [{\rm GeV}^2]  \nonumber\\
& \chi^2=5.86\cdot 10^{-2} .
\end{eqnarray}
Our best fit of the logarithmic form,
\begin{eqnarray}
\alpha_{\rm log}(0)& = &-10.89 \pm 7.59 
, \nonumber\\ 
\Big( \gamma \nu \Big)_{\rm log}& = &5.0 \pm 3.92, \nonumber\\
T_{\rm log }& = &130.74\pm 5.39 \  [{\rm GeV}^2] \nonumber\\
&  \chi^2=1.45\cdot 10^{-2} .
\end{eqnarray}
is shown in Fig. 1(b). In both cases, the large errors on the extracted 
parameters reflect the fact that there are only 8 data points for the 
trajectory. 

The two fits are indistinguishable in the region of the fit. In our case, 
since we {\it know} that in our model there is only a finite number of bound 
states, the square-root form (which terminates in general, and in our fit its 
termination point is near the termination of our bound-state data) is favored 
over the logarithmic form (which does not terminate). However, in practise, 
when one has no means to recognize whether the set of data is a part of a 
finite or an infinite ensemble, it is impossible to distinguish between the 
two forms based on fits to the bound state data.

In view of this, we extend our considerations on how to numerically determine 
the form of a Regge trajectory from data. There is more information about 
Regge trajectories than just the masses of the bound states. Total 
cross sections can be related to the value of the relevant Regge trajectory at 
the origin (i.e., the intercept of the trajectory). In the region of negative 
$t,$ the value of $\alpha(t)$ can be determined from the relevant differential
cross sections. If there are reliable data at large\footnote{Unfortunately, 
according to our numerical simulations, data at larger negative $t$ values, 
and with greater measurement precision, are both needed.} negative
$t$ in addition to bound states, it might be possible to obtain a more 
stringent constraint on the form of trajectory by fitting both sets of data 
simulataneously, providing the form is universally valid for all $t$.


\section{Analytic model}
Our model calculation for heavy quarks revealed that the Regge trajectory 
formed by the bound states is certainly nonlinear but can be equally well 
fitted by a square-root form or a logarithmic form.

In this section we try to shed some light on this observation using  
analytic model for a massless string with variable tension. The ends of the 
string can be massive. The model is intended to mimic a flux tube stretched 
between two quarks, and the varrying tension is to reflect possible dynamical 
effects such as weekening of the flux tube due to pair creation.

The particular merit of the model  for our purposes is that, given the form of
a Regge trajectory, we are able to recover the form of underlying potential, 
in many cases in analytic form. We will show in this model why the two 
forms of Regge trajectories (i.e. square-root and logarithmic) are both likely
to well approximate the Regge trajectory formed by the bound states of the 
unquenched lattice QCD potential (4).

This section is organized as follows: first we review the standard 
relativistic Nambu-Goto string with massive ends and show that in the 
nonrelativistic limit this model simply corresponds to a rigid rod  with a 
linear potential generated between its massive ends. After introducing the 
generalized string model, we show that it corresponds in the nonrelativistic 
limit to a rigid rod with arbitrary potential. Details fo the dynamics of the 
generalized string with massive ends, such as derivation of expressions for 
its energy and orbital angular  momentum, can be found in the Appendix B.
The key part of this section is devoted to the generalized string with 
massless ends. Within this framework we find potentials that lead to 
square-root and logarithmic trajectories, respectively. Since the dynamics of 
mesons (such as the termination of Regge trajectories) is dominated by the 
behavior of the flux tube, we claim that what we learn from the string with 
massless ends is qualitatively relevant to the case of massive quarks as well. 

\subsection{The Nambu-Goto string and the generalized string.}
The action of the standard relativistic Nambu-Goto string with massive ends 
in the parametrization $\tau =t=x^0$ is written as $(\sigma $ is the string 
tension) \cite{BN2}
\begin{eqnarray}
S=-\sigma \int _{t_1}^{t_2}dt\int _0^\pi ds  \sqrt{x'^2(1-\dot{x}^2)+
(\dot{x}x')^2}-\sum_{i=1,2}m_i\int _{t_1}^{t_2}dt\;\!\sqrt{1-\dot{x}_i^2},
\end{eqnarray}
$$x\equiv {\bf x}={\bf x}(t,s  ),\;\;\;x_i\equiv x(t,s  _i),
\;\;i=1,2,\;\;s  _1=0,\;\;s  _2=\pi ,$$
where from now on the dot and the prime stand for the derivative with respect 
to $t$ and $s  ,$ respectively, unless otherwise specified.

To illuminate the physical meaning of this model, let us consider for the 
moment its nonrelativistic limit. 
In the nonrelativistic limit, $|\dot{x}(t,s  )|\ll 1,$ $|\dot{x}_i|\ll 1,$
and the action reduces to \cite{BN2}
\begin{eqnarray}
S=-\sigma \int _{t_1}^{t_2}dt\int _0^\pi ds  \sqrt{x'^2}-\sum_{i=1,2}m_i
\int _{t_1}^{t_2}dt+\sum _{i=1,2}\frac{m_i}{2}\int _{t_1}^{t_2}dt\;\!\dot{
x}_i^2.
\end{eqnarray}
Integration over $s  $ gives the length of the string (under the assumption 
that there are no singularities on the string). The variation of the first 
term in the nonrelativistic action, Eq. (11),  with respect to the string 
coordinates leads to the requirement on the
string to have the form of a linear rod connecting the massive ends. The
effective action that leads to the equations of motion of the massive ends
is therefore
\begin{eqnarray}
S_{eff}=\int _{t_1}^{t_2}dt\left( -\sigma |x_1(t)-x_2(t)|+
\sum _{i=1,2}\frac{m_i\dot{x}_i^2}{2}\right) .
\end{eqnarray}
Hence, in the nonrelativistic limit, the string generates a linear potential
between its massive ends: $V(|x_1-x_2|)=\sigma |x_1-x_2|.$

Here we generalize the standard string formulation described above to
the case of an arbitrary potential between the string massive ends. Such 
generalization is done by the modification of the standard, constant, string 
tension into an  effective string tension which is a function of $|x|,$ as 
follows:
\begin{eqnarray}
S_{gen}=-\int _{t_1}^{t_2}dt\int _0^\pi ds  \;\!\sigma (|x|)\sqrt{
x'^2(1-\dot{x}^2)+(\dot{x}x')^2}-\sum_{i=1,2}m_i\int _{t_1}^{t_2}dt\;\!
\sqrt{1-\dot{x}_i^2}.
\end{eqnarray}
The action is similar to that of the standard Nambu-Goto string. In the 
nonrelativistic limit, however, in place of (13) one will now obtain
$$S_{gen,eff}=\int _{t_1}^{t_2}dt\left( -\int _0^\pi ds  \;\!
\sigma (|x|)|x'|+\sum _{i=1,2}\frac{m_i\dot{x}_i^2}{2}\right) $$
\begin{eqnarray}
=\int _{t_1}^{t_2}dt\left( -V(|x_1-x_2|)+\sum _{i=1,2}\frac{m_i\dot{
x}_i^2}{2}\right) .
\end{eqnarray}
In contrast to the previous case of the standard string, it is seen in the
above relations that now
\begin{eqnarray}
\sigma (|x|)=\frac{1}{|x'|}\;\!\frac{dV(|x|)}{ds  }=\frac{dV(|x|)}{d|x|},
\end{eqnarray}
i.e., the effective string tension is the derivative of a potential with
respect to the distance. Obviously, in the case of a linear potential, the
effective string tension reduces to the standard (constant) one. 

Similarly to the standard case of a constant string tension which represents 
the relativization {\it \`{a} la} Poincar\'{e} of the nonrelativistic two-body 
problem with linear potential \cite{Shav}, the generalized string can be 
considered as the relativization of a nonrelativistic two-body problem with 
an arbitrary potential. Details of the dynamics of the generalized string 
model are given in Appendix B.  

\subsection{Generalized massless string}
Since we are interested in dynamical issues related to the properties of the 
flux tube, without a loss of generality it is sufficient for us to consider 
the case of a generalized string with a massless ends. This case is obviously
more tractable than the case with arbitrary massess, and nevertheless exhibits
the same qualitative features. 

The energy and orbital momentum of the generalized massless string, 
$m_1=m_2=0,$ is given by
\begin{eqnarray}
E=2\int _0^R\frac{d\rho \;\!\sigma (\rho )}{\sqrt{1-\omega ^2\rho ^2}},\;\;\;
J=2\int _0^R\frac{d\rho \;\!\sigma (\rho )\omega \rho^2}{\sqrt{1-\omega ^2\rho
^2}}, 
\end{eqnarray}
where $R=1/\omega $ is half of the string length for a given $\omega .$ The
condition $\omega R=1$ follows from, e.g., Eqs. (B.14) with 
$m_i\rightarrow 0.$ 

By eliminating $\omega $ from Eqs. (17) one can obtain $J$ as a function of 
$E^2,$ the Regge trajectory. It will be shown elsewhere \cite{prep} that it 
is possible to uniquely recover the potential $(V(r)\sim \int d\rho $ $\sigma 
(\rho ))$ from the known analytic form of Regge trajectory, for both massless 
and massive generalized strings, and the techniques of the corresponding 
inverse problem will be presented in detail. In this way, we have recovered 
the potentials that correspond to both the square-root and logarithmic Regge 
trajectories. For our present purposes, here we present only the final results.

\subsubsection{Square-root trajectory}
The potential $(\sigma ,\mu ={\rm const,}$ $V(\rho )\rightarrow \sigma /2\mu $
as $\rho \rightarrow \infty ,$ and hence $E\rightarrow \sigma /\mu )$
\begin{eqnarray}
V(\rho )=\frac{\sigma }{\pi \mu }\arctan (\pi \mu \rho ),
\end{eqnarray}
for which
\begin{eqnarray}
\sigma (\rho )=\frac{dV(\rho )}{d\rho }=\frac{\sigma }{1+(\pi \mu \rho )^2},
\end{eqnarray}
leads, via (17), to
\begin{eqnarray}
E=\frac{\pi \sigma }{\sqrt{\omega ^2+\pi ^2\mu ^2}},\;\;\;J=\frac{\sigma }{
\pi \mu ^2}\left( 1-\frac{\omega }{\sqrt{\omega ^2+\pi ^2\mu ^2}}\right) .
\end{eqnarray}
Eliminating $\omega $ from the above relations gives
\begin{eqnarray}
J=\frac{1}{\pi \mu }\left( \sigma /\mu -\sqrt{(\sigma /\mu )^2-E^2}\right) ,
\end{eqnarray}
i.e., the square-root Regge trajectory. For $E\ll \sigma /\mu ,$ it reduces
to an (approximate) linear trajectory, $J\simeq E^2/(2\pi \sigma ).$ 
\begin{figure}
\centerline{\psfig{file=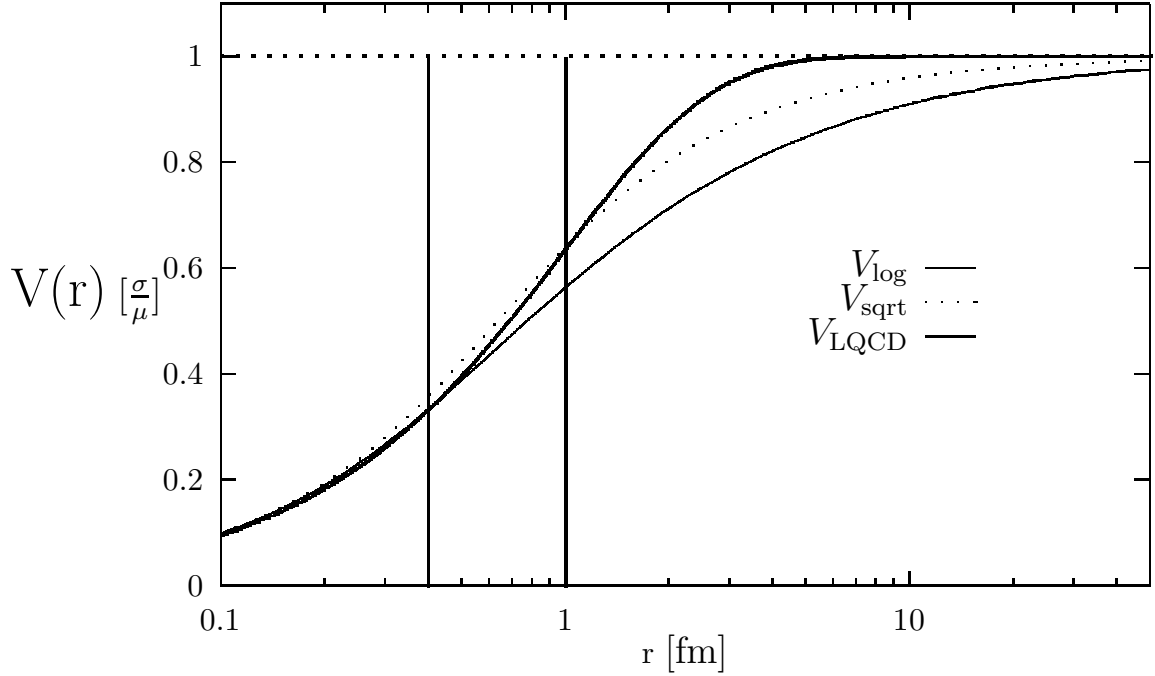,width=6in}}
\caption[]{Potentials which in our analytical model lead to logarithmic and 
square-root Regge trajectories, respectively, compared to the unquenched 
lattice QCD potential used in our heavy-quarkonia calculation.}
\end{figure}

\subsubsection{Logarithmic trajectory}
The potential
\begin{eqnarray}
V(\rho )=\frac{\sigma }{2\pi \mu }\left( 2\arctan (2\pi \mu \rho )-\frac{
\log [1+(2\pi \mu \rho )^2]}{2\pi \mu \rho }\right) ,
\end{eqnarray}
for which 
\begin{eqnarray}
\sigma (\rho )=\sigma \;\!\frac{\log [1+(2\pi \mu \rho )^2]}{
(2\pi \mu \rho )^2},
\end{eqnarray}
leads to
\begin{eqnarray}
E=\frac{\sigma }{2\pi \mu ^2}\left( \sqrt{\omega ^2+4\pi ^2\mu ^2}-\omega 
\right),\;\;\;J=\frac{\sigma }{2\pi \mu ^2}\;\!\log \frac{\omega +\sqrt{
\omega ^2+4\pi ^2\mu ^2}}{2\omega },
\end{eqnarray}
from which eliminating $\omega $ (viz., $\omega =\pi (\sigma ^2-\mu ^2E^2)/(
\sigma E))$ gives
\begin{eqnarray}
J=-\frac{\sigma }{2\pi \mu ^2}\;\!\log \left( 1-\frac{E^2}{(\sigma /\mu )^2}
\right) ,
\end{eqnarray}
i.e., the logarithmic Regge trajectory. For $E\ll \sigma /\mu ,$ it again
reduces to an (approximate) linear form, $J\simeq E^2/(2\pi \sigma ).$ 

Fig. 2 shows both  potentials together with the confining part of the 
screened lattice QCD potential that we used in the heavy quark model in the 
previous section. All potentials are normalised to the same asymptotic value. 
The vertical lines show the region of  the  distance $r$ which is relevant for
the bound states.

All three potentials are very close in the region of bound state physics. 
Therefore, one can conclude that they would lead to similar Regge trajectories
as far as the bound states not close to the trajectory thresholds are 
concerned. One could 
further speculate that since the lattice potential is steeper than the 
square-root potential (which in turn, is steeper than the logarithmic 
potential) in asymptotia, it is reasonable to expect that the bound states of 
the lattice potential terminate at a lower ${\ell}$ than the square-root 
trajectory, and that the square-root trajectory approximates the lattice 
potential trajectory better 
for states near the trajectory threshold than the logarithmic trajectory. 
(Recall that the real part of the logarithmic trajectory does not terminate.)

Combining the conclusions of the analytic model for the generalized string 
with those of the heavy quark model of the previous section, we conclude that 
even though both nonlinear forms considered may appear to be a good 
approximation to QCD, the square-root form is likely to be more realistic. 
Therefore, we use the square-root form for phenomenological purposes. 

\section{Trajectory parameters and spectroscopy}
In this section we determine, assuming that Regge trajectories are of the form 
(8), trajectories thresholds and intercepts using various 
experimental information. Typically, we use masses of a few known lowest lying
states, and in the case the $\rho$ trajectory we also use the value of the 
intercept (which is known and well-established) found from exchange processes.
The value of $\gamma$ (the universal asymptotic slope) is fit to the $\rho$ 
trajectory, and then taken as universal for all other trajectories.

Whenever possible, we try to use as inputs states that are believed to be pure 
quark-antiquark states with definite flavors. For example, we do not use the 
masses of the observed $\phi$, $f$ or $\eta (\eta')$ states to find the 
parameters of $s\bar{s}$ trajectory because the physical states are mixtures 
of $s\bar{s}$ with the light quark-antiquark components. 

In some cases, e.g. for tensor mesons and axial-vector mesons, there are not 
enough data to determine the parameters. We are forced to make aditional 
assumptions. In particular, we assume that the thresholds of parity partner 
trajectories coincide. We comment further on this assumption where it is 
introduced in Section 4.2. Here we wish to mention that 
our additional assumptions are justified {\it a posteriori} by our 
excellent results.

The approach has more predictive power than one would naively expect. This is 
because the parameters for different flavors are related by 

(i) additivity of intercepts, 
\begin{eqnarray}
\alpha _{i\bar{i}}(0)+\alpha _{j\bar{j}}(0)=2\alpha _{j\bar{i}}(0),
\end{eqnarray}
where $i,j$($\bar{i},\bar{j}$) refer to the quark (antiquark)
flavor\footnote{This additivity is a firmly established theoretical
constraint on Regge trajectories (for an extensive list of references
see \cite{BG}).}, and

(ii) additivity of inverse slopes near the origin,
\begin{eqnarray}
\frac{1}{\alpha'_{i\bar{i}}}+\frac{1}{\alpha'_{j\bar{j}}}= 
\frac{2}{\alpha'_{j\bar{i}}},
\end{eqnarray}
which is favored over another constraint suggested in the literature, 
factorization of slopes, $\alpha'_{i\bar{i}}\alpha'_{j\bar{j}}=(\alpha'_{
j\bar{i}})^2,$ by the heavy quark limit \cite{BG}. These two additivity 
requirements are independent of which specific form is assumed for the 
trajectories.

Once the parameters of the square-root trajectories are known, the first 
obvious application is meson spectroscopy. We calculate masses of a few 
excited states lying on each of the trajectories under consideration, and 
compare with experimental data from ref. \cite{pdg}. Where there are no data 
available, our results are predictions.

We start with vector meson trajectories, and then consider tensor meson, 
pseudoscalar  meson, and finally, axial-vector meson trajectories. The section 
concludes with a brief discussion of the results.

\subsection{Vector mesons}
We start with the $\rho $ trajectory. 
The intercept of this trajectory is well established. From the behavior of
the differential cross section of the process $\pi ^{-}p\rightarrow \pi ^0n$
the intercept $\alpha _\rho (0)$ has been extracted to be 0.58 \cite{LS}, 
$0.56\pm 0.01$ 
\cite{AC}, $0.56\pm 0.02$ \cite{Bol}, 0.53 \cite{Bar}. Extensive analysis of 
this process by H\"{o}hler {\it et al.} \cite{Hoh} leads to $\alpha _\rho (0)=
0.55$ \cite{HJ}. From the difference of the total cross sections of $\pi ^{
+}p$ and $\pi ^{-}p$ scattering $\alpha _\rho (0)$ has been inferred to be 
$0.57\pm 0.01$ \cite{Hen}, $0.55\pm 0.03$ \cite{Car}. Bouquet finds from the 
dual topological unitarization \cite{Bou} $0.51\leq \alpha _\rho (0)\leq 
0.54.$ All of the above values are consistent with  
\begin{eqnarray}
\alpha _\rho (0)=0.55, 
\end{eqnarray}
in agreement with the value of the $\rho $ trajectory intercept extracted by 
Donnachie and Landshoff from the analysis of $pp$ and $\bar{p}p$ scattering 
data in a simple pole exchange model \cite{DL}. We therefore take the value of 
the intercept, Eq. (28) as 
one of the constraints on the 
form of the $\rho $ trajectory.
Two more constraints that are needed are provided by the mass and spin of 
$\rho$ and $\rho_3$, i.e.:
(i) $\alpha _\rho (M^2_\rho )=1,$ (ii) $\alpha _\rho (M^2_{\rho _3})=3,$ where
$M_\rho =769.0 \pm 0.9$ MeV and $M_{\rho _3}=1688.8\pm 2.1$ MeV \cite{pdg}.

Inserting these values into the functional form (8), we extract the parameter 
values
\begin{eqnarray}
\gamma =3.65\pm 0.05\;{\rm GeV}^{-1},\;\;\;\sqrt{T_\rho }=2.46 \pm 0.03\;
{\rm GeV.}
\end{eqnarray}
It is interesting to note that the values of parameters we find based on 
spectroscopy and the known intercept  are in excellent 
agreement with 
\begin{eqnarray}
\gamma =3.72\pm 0.30\;{\rm GeV}^{-1},\;\;\;\sqrt{T_\rho }=2.50\pm 0.10\;
{\rm GeV,}
\end{eqnarray}
extracted in ref. \cite{KT} for a similar form  of the $\rho $ trajectory 
from the analysis of $\pi N$ charge-exchange scattering data.\footnote{The
$\rho $ trajectory adopted in \cite{KT} contains an additional term
$-0.14\sqrt{4M_\pi ^2-t},$ to take into account nonzero resonance widths, 
which reduces the planar intercept $\alpha_{\rho}(0)$ down to 0.51.} Earlier 
analysis of ref.
\cite{KP} found $\sqrt{T_\rho }=2.4\pm 0.4$ GeV.  

The parameters of the $K^\ast $ trajectory are obtained by  using $\alpha 
_{K^\ast }(M^2_{
K^\ast })=1$ with $M_{K^{\ast 0}}=896.1\pm 0.3$ MeV  and $\alpha _{K^\ast 
}(M^2_{
K_3^\ast })=3$ with $M_{K_3^{\ast 0}}=1776\pm 7$ MeV \cite{pdg}, and taking 
the value of $\gamma $ found from $\rho$ spectra as a universal slope in 
asymptopia. This yields
\begin{eqnarray}
\sqrt{T_{K^\ast }}=2.58\pm 0.03\;{\rm GeV.} \nonumber\\
\alpha_{K^\ast }(0)=0.414\pm 0.006
\end{eqnarray}
The value of the intercept is in an excellent agreement with
the results of the analysis of hypercharge 
exchange processes $\pi ^{+}p\rightarrow K^{+}\Sigma ^{+}$ and $K^{-}p
\rightarrow \pi ^{-}\Sigma ^{+}$ \cite{VKKT}.

Taking the parameters of the $\rho $ and $K^\ast $ trajectories as known, 
those for the $\phi $ trajectory may be  obtained from the requirements of 
additivity of inverse slopes and intercepts, Eqs. (26),(27):
\begin{eqnarray}
\alpha _\phi (0)=0.25\pm 0.02,\;\;\; \nonumber \\
\sqrt{T_\phi }=2.59\pm 0.11\;{\rm GeV.}
\end{eqnarray}

Similarly, one can obtain the parameters of trajectories for the states 
containing $c$- and $b$-quarks, using the value of the universal slope 
$\gamma$ found from $\rho$-trajectory and masses of the corresponding states. 
In particular, we use masses of $D^{\ast}$, $J/\psi$, $B^{\ast}$ and 
$\Upsilon$ as inputs. The remaining trajectories are determined from the 
requirements of additivity of inverse slopes and of additivity of intercepts, 
Eqs. (26),(27). The parameters of these vector meson trajectories are 
summarized in Table I.

\begin{center}
\hspace*{-1cm}
{\footnotesize
\begin{tabular}{|c|c|c|c|c|} \hline
  & $\rho $ & $K^\ast $ & $\phi $ & 
\\ \hline 
$\alpha (0)$ & $0.55$ & $0.414 \pm 0.006$ & $0.27\pm 0.01$ & 
 \\ \hline
$\sqrt{T},$ GeV & $2.46\pm 0.03$ & $2.58\pm 0.03$ & $2.70\pm 0.07$ & 
    \\ \hline
       & $D^\ast $ & $D_s^\ast $ & $J/\psi $ &  
\\ \hline
$\alpha (0)$ & $-1.02\pm 0.05$ & $-1.16\pm 0.05$ & $-2.60\pm 0.10$ & 
  \\ \hline
$\sqrt{T},$ GeV & $3.91\pm 0.02$ & $4.03\pm 0.04$ & $5.36\pm 0.05$ & 
    \\ \hline
 & $B^\ast $ & $B_s^\ast $ & $B_c^\ast $ & $\Upsilon $     \\ \hline
 $\alpha (0)$ 
 &$-7.13\pm 0.17$ & $-7.27\pm 0.17$ & $-8.70\pm 0.18$ & $-14.81\pm 0.35$  \\  
\hline
 $\sqrt{T},$ GeV &$7.48\pm 0.02$ & $7.60\pm 0.04$ & $8.93\pm 0.03$ & $12.50\pm 
0.02$ 
  \\ \hline
\end{tabular}
}
\vskip0.2in
\end{center}
{\bf Table I.} Parameters of the vector meson trajectories of the form (8). 
The intercept of the $\rho$ trajectory was taken as an input.
 \\

Using the parameters shown in Table I, we calculate masses of 
the spin-1, spin-3 and spin-5 states lying on these trajectories. Our results 
are compared with data from \cite{pdg} in Table II. In this, and subsequent 
Tables IV, VI, VIII, the values used as input for our analysis are shown in
boldface. The masses of the states for which there are no data available 
should be considered as our predictions.

\begin{center}
\begin{tabular}{|c|cc|cc|cc|} \hline
 & \multicolumn{2}{c|}{$J=1$} & \multicolumn{2}{c|}{$J=3$} &
 \multicolumn{2}{c|}{$J=5$}   \\ \hline
 & This work & ref. \cite{pdg} & This work & ref. \cite{pdg} & This work &
ref. \cite{pdg}   \\ \hline
$\alpha _\rho (t)$ & ${\bf 769.0\pm 0.9}$ & $769.0\pm 0.9$ & ${\bf 1688.8\pm 
2.1}$ & $1688.8\pm 2.1$ & $2124\pm 19$ &       \\ \hline
$\alpha _{K^\ast }(t)$ & ${\bf 896.1\pm 0.3}$ & $896.1\pm 0.3$ & ${\bf 1776
\pm 7}$ & $1776\pm 7$ & $2215\pm 21$ &       \\ \hline
$\alpha _\phi (t)$ & $1015\pm 17$ & 1019.4 & $1863\pm 31$ & $1854\pm 7$ & 
$2305\pm 42$ &       \\ \hline
$\alpha _{D^\ast }(t)$ & ${\bf 2006.7\pm 0.5}$ & $2006.7\pm 0.5$ & $2721\pm
23$ &   & $3191\pm 22$ &       \\ \hline
$\alpha _{D_s^\ast }(t)$ & $2102\pm 29$ & $2106.6\pm 2.1\pm 2.7$ & $2808\pm 
28$ &   & $3279\pm 30$ &       \\ \hline
$\alpha _{J/\psi }(t)$ & {\bf 3096.9} & 3096.9 & $3753\pm 41$ &   & $4240\pm
39$ &       \\ \hline
$\alpha _{B^\ast }(t)$ & ${\bf 5324.9\pm 1.8}$ & $5324.9\pm 1.8$ & $5814\pm
51$ &   & $6217\pm 46$ &       \\ \hline
$\alpha _{B_s^\ast }(t)$ & $5411\pm 58$ & $5416.3\pm 3.3$ & $5901\pm 53$ &  
 & $6306\pm 49$ &       \\ \hline
$\alpha _{B_c^\ast }(t)$ & $6356\pm 80$ &   & $6853\pm 72$ &   & $7276\pm 65$
 &       \\ \hline
$\alpha _\Upsilon (t)$ & ${\bf 9460.4\pm 0.2}$ & $9460.4\pm 0.2$ & $9906\pm 
91$ &   & $10304\pm 84$ &       \\ \hline
\end{tabular}
\end{center}
{\bf Table II.} Comparison of the masses of the spin-1, spin-3 and spin-5
states given by ten vector meson trajectories of the form (8) with data.
All masses are in MeV.
 \\

\subsection{Tensor meson trajectories} 
Tensor meson trajectories are parity partners of the vector meson 
trajectories. In a nonrelativistic theory, tensor meson trajectories are 
degenerate with vector meson trajectories. In a field theory, both acquire 
different corrections to the intercept, but they can still be expected nearly 
degenerate in the bound state region. From the form of the trajectory, Eq. 
(8), and since the intercept is in practise much smaller than the other terms 
in Eq. (8), it is clear that trajectories can be near-degenerate in the 
bound-state region even if their intercepts differ by a large percentage, but 
not if their thresholds are very different. For this reason, and to reduce the
number of free parameters, we {\it assume} that the thresholds of tensor meson
trajectories are the same as the thresholds of the vector meson ones.

Our assumption is supported by analysis of the processes  $\pi ^{+}p
\rightarrow K^{+}\Sigma ^{+}$ and $K^{-}p\rightarrow \pi ^{-}\Sigma ^{+}$ 
\cite{VKKT}. Most of the analysed experiments 
show the same slope at the origin for $K^{\ast}$ and $K_2^{\ast}$ 
trajectories. In our formalism, this translates to equal thresholds.

To further  test this assumption, we first fit the threshold of the 
$K_2^\ast $ trajectory which is the only one with more than one state. 
(A recent measurement of the $a_4$ mass \cite{VES1,VES2} is four standard 
deviations below the previous value \cite{pdg}, see discussion in the text 
below.) The parameters of the $K_2^\ast $ 
trajectory are fixed by using $\gamma $ determined for $\rho$ trajectory, 
$\alpha _{K_2^\ast }(M^2_{K_2^\ast })=2$ with $M_{K_2^{\ast 0}}=1432.4\pm 1.0$
MeV and $\alpha _{K_2^\ast }(M^2_{K_4^\ast })=2$ with $M_{K_4^{\ast 0}}=2045
\pm 9$ MeV \cite{pdg}. The calculation yields
\begin{eqnarray}
\sqrt{T_{K_2^\ast }} & = &2.64\pm 0.03\;{\rm GeV} 
\end{eqnarray}
The value of the $K_2^\ast $ trajectory threshold is consistent 
with that of the $K^\ast $ trajectory given in Table I, supporting our 
assumption. Moreover, it is in excellent agreement with the values for the 
slope shown in ref. \cite{VKKT}.


We therefore conclude that our assumption that the thresholds of the parity 
partners are the same is plausible. Under this simplifying assumption
only one state on each trajectory is needed to completely fix the parameters. 
We choose to fit $a_2$, $K_2^*$, $\chi_{c 2}$ and $\chi_{b 2}$ trajectories 
using the masses of the corresponding lowest lying state. The remaining 
trajectories are determined using the addititivity requirements.
Parameters found in this way are presented in Table III.

\begin{center}
\hspace*{-1cm}
{\footnotesize
\begin{tabular}{|c|c|c|c|c|} \hline
  & $a_2 $ & $K_2^\ast $ & ${f_2^{'}} $ & 
\\ \hline 
$\alpha (0)$ & $0.60 \pm 0.03$ & $0.42 \pm 0.03$ & $0.22 \pm 0.07$ & 
 \\ \hline
       & $D_2^\ast $ & $D_{s2}^\ast $ & $\chi _{c2}(1P) $ & 
\\ \hline
$\alpha (0)$ & $-1.16 \pm 0.05$ & $-1.35 \pm 0.05$ & $-2.93 \pm 0.09$ & 
 \\ \hline
 & $B_2^\ast $ & $B_{s2}^\ast $ & $B_{c2}^\ast $ & $\chi _{b2}(1P)$     \\ 
\hline
 $\alpha (0)$ 
 &$-7.62 \pm 0.13$ & $-7.81 \pm 0.13$ & $-9.39 \pm 0.14$ & $-15.85 \pm 0.26$ 
 \\  \hline
\end{tabular}
}
\vskip0.2in
{\bf Table III.} Parameters of tensor meson trajectories of the form (8).
 \\
\end{center}

As before, we use the knowledge of Regge trajectories  for spectroscopy 
purposes. Our results are compared with data from \cite{pdg} in Table IV. The 
masses of the states for which there are no data available should again be 
considered  our predictions.

\begin{center}
\begin{tabular}{|c|cc|cc|cc|} \hline
 & \multicolumn{2}{c|}{$J=2$} & \multicolumn{2}{c|}{$J=4$} &
 \multicolumn{2}{c|}{$J=6$}   \\ \hline
 & This work & ref. \cite{pdg} & This work & ref. \cite{pdg} & This work &
ref. \cite{pdg}   \\ \hline
$\alpha _{a_2} (t)$ & ${\bf 1318.1\pm 0.6}$ & $1318.1\pm 0.6$ & $1927\pm 18$ &
 $2020\pm 16$ & $2256\pm 21$ & $2450\pm 130$   \\ \hline
$\alpha _{K_2^\ast }(t)$ & ${\bf 1432.4\pm 1.3}$ & $1432.4\pm 1.3$ & $2026\pm
20$ & $2045\pm 9$ & $2357\pm 24$ &       \\ \hline
$\alpha _{f'_2}(t)$ & $1544\pm 37$ & $1525\pm 5$ & $2124\pm 40$ &   & $2457\pm
48$ &       \\ \hline
$\alpha _{D_2^\ast }(t)$ & $2454\pm 23$ & $2458.9\pm 2.0$ & $3010\pm 22$ &   &
$3390\pm 21$ &       \\ \hline
$\alpha _{D_{s2}^\ast }(t)$ & $2560\pm 27$ & $2573.5\pm 1.7$ & $3109\pm 29$ &
 & $3489\pm 31$ &       \\ \hline
$\alpha _{\chi _{c2}(1P)}(t)$ & ${\bf 3556.2\pm 0.1}$ & $3556.2\pm 0.1$ & 
$4092\pm 39$ &   & $4498\pm 38$ &       \\ \hline
$\alpha _{B_2^\ast }(t)$ & $5698\pm 45$ & $5698\pm 12$ & $6122\pm 42$ &   & 
$6472\pm 39$ &       \\ \hline
$\alpha _{B_{s2}^\ast }(t)$ & $5797\pm 47$ &   & $6220\pm 45$ &   & $6570\pm
43$ &       \\ \hline
$\alpha _{B_{c2}^\ast }(t)$ & $6780\pm 52$ &   & $7213\pm 50$ &   & $7582\pm 
49$ &       \\ \hline
$\alpha _{\chi _{b2}(1P)}(t)$ & ${\bf 9913.2\pm 0.6}$ & $9913.2\pm 0.6$ & 
$10310\pm 72$ &   & $10665\pm 68$ &       \\ \hline
\end{tabular}
\end{center}
{\bf Table IV.} Comparison of the masses of the spin-2, spin-4 and spin-6
states given by ten tensor meson trajectories of the form (8) with data.
All masses are in MeV. See the text regarding the $a_4$ and $a_6$ masses.
 \\

Note that all the masses agree very well for the available data, except for
$M_{a_4}.$ With respect to this, we wish to mention that recent analyses by 
the VES collaboration of the reactions $\pi ^{-}Be\rightarrow \pi ^{+}2\pi ^{-}
2\pi ^{0}Be$ \cite{VES1} and $\pi ^{-}Be\rightarrow \pi ^{+}2\pi ^{-}2Be$ 
\cite{VES2} reveal the mass of the $a_4$ resonance as seen in the 
$a_4\rightarrow \omega \rho $ and $a_4\rightarrow f_2\pi $ decays, 
respectively: $1944\pm 8\pm 50$ MeV \cite{VES1} and $1950\pm 20$ MeV 
\cite{VES2}. Our value in Table IV is in excellent agreement with these latest 
measurements.

According to our fitted square-root form for the $a_2$ trajectory, the last 
state below threshold is $J=8$. This conclusion is very sensitive to the 
functional form assumed, even though the threshold value itself is not (see 
Section 2). Generally, as the mass approaches the threshold, we expect to find 
larger discrepancies between the observed states and predictions of any
specific form of trajectory, Eq. (5). Nonetheless, the discrepancy for the 
next-to-last, $J=6$ state appears to be less than 10\%. Note also that our 
$K_4^\ast $ prediction appears to be below the data, consistent with $a_4$ and
$a_6$ deviations, suggesting a possible systematic effect. This raises the 
possibility that the growth of the square-root form of trajectory near the 
termination point is too rapid and that the true trajectories are somewhat 
flatter in this region. Thus, it is possible that the $J=6$ state is actually 
the last state on these trajectories. This point of view is further supported 
by the large width of the $a_6$, as would be expected for a state near the 
termination point. Hence we also expect the $K_6^\ast $ to be similarly broad.

\subsection{Pseudoscalar meson trajectories}
Here we calculate the parameters of the pseudoscalar meson trajectories.  We 
start with the pion trajectory and fix its intercept and threshold by using the
masses of the two lowest lying states, $M_{\pi ^0}=135$ MeV and $M_{\pi _2}=
1677\pm 8$ MeV \cite{pdg}, in relations $\alpha _\pi (M^2_\pi )=0,$ $\alpha _
\pi (M^2_{\pi _2})=2,$ and $\gamma $ from Eq. (29). 

Similarly, the intercept and the threshold of the $K$-trajectory are 
determined using the massses of $K$ and $K_2$.

The parameters of charmed mesons trajectories are found using the masses of 
$D$ and $\eta_c$, utilizing the additivity requirements, Eqs. (26),(27), 
and the parameters of the light mesons ($\pi$-trajectory) found above.

If one tries to procceed similarly in the case of mesons containing the $b$ 
quark, and use the only two experimentally known masses as inputs (the mass of 
$B$ and $B_s$) in conjunction with the parameters of $\pi$- and 
$K$-trajectory, the error estimates on the extracted values exceed two hundred
percent. The results turn out to be highly unstable with respect to the 
threshold of the $\pi $-trajectory. This is a peculiar consequence of the 
pion intercept being close to zero, and $J=0$ of the lowest lying state.

Conversely, this observation allows us to convert the problem into an 
additional self-consistency check. We use the $b$-quark meson spectra to 
extract information about the $\pi$ trajectory. To do that, an additional 
input is needed. We use the mass of $\eta_b$ given in ref. \cite{NRQCD}, and 
fit the value of the pion threshold in addition to the parameters of 
$b$-containing mesons. Recall that the light meson parameters 
affect the heavy quark parameters only through the additivity requirements. 
Since the additivity requirements are well established, such a  
self-consistency check is quite nontrivial. 

The parameters of the pseudoscalar trajectories that we obtain are shown in 
Table V. The parameters of the $\pi$-trajectory given in the Table come from 
fitting the pion spectra. From the bottom spectra we obtain the following 
value of the pion threshold:
\begin{eqnarray}
\sqrt{T_\pi }= 2.79 \pm  0.40 \, \, \, {\rm GeV}
\end{eqnarray}
Note the agreement with the value given in Table V.

\begin{center}
\hspace*{-1cm}
{\footnotesize
\begin{tabular}{|c|c|c|c|c|} \hline
  & $\pi $ & $K $ & $\eta_s $ & 
\\ \hline 
$\alpha (0)$ & $-0.0118 \pm 0.0001$ & $-0.151 \pm 0.001$ & $-0.291\pm 0.003$ & 
 \\ \hline
$\sqrt{T},$ GeV & $2.82\pm 0.05$ & $2.96\pm 0.05$ & $3.10\pm 0.11$ & 
    \\ \hline
       & $D $ & $D_s $ & $\eta_c $ &  \\ \hline
$\alpha (0)$ & $-1.61105\pm 0.00005$ & $-1.751\pm 0.001$ & $-3.2103 \pm 
0.0001$ &   \\ \hline
$\sqrt{T},$ GeV & $4.16\pm 0.03$ & $4.29\pm 0.06$ & $5.49\pm 0.02$ & 
    \\ \hline
 & $B $ & $B_s $ & $B_c$ & $\eta_c $     \\ \hline
 $\alpha (0)$ 
 &$-7.41\pm 0.17$ & $-7.54\pm 0.17$ & $9.00\pm 0.17$ & $-14.80\pm 0.34$  \\  
\hline
 $\sqrt{T},$ GeV &$7.89\pm 0.16$ & $8.01\pm 0.16$ & $9.24\pm 0.12$ & $12.98\pm 
0.24$   \\ \hline
\end{tabular}
}
\vskip0.2in
{\bf Table V.} Parameters of the pseudoscalar meson trajectories of the form 
(8).\footnote{The parameters for the $K$-trajectory were found using the mass 
of $K_2$ from \cite{pdg}. If we instead use a mass of the corresponding 
{\it pure} $n\bar{s}$ state as found in ref. \cite{Dwave}, i.e. $M_{K_2}=
1762 \pm 18$ GeV, the parameters change slightly: the intercept $-0.153 \pm
 0.003$, and the threshold $2.93 \pm 0.07$ GeV.}
 \\
\end{center} 
 
Our results for pseudoscalar spectroscopy are compared with data from 
\cite{pdg} in Table VI. The masses of the states for which there are no data 
available should once again be considered  our predictions.

\begin{center}
\begin{tabular}{|c|cc|cc|cc|} \hline
 & \multicolumn{2}{c|}{$J=0$} & \multicolumn{2}{c|}{$J=2$} &
 \multicolumn{2}{c|}{$J=4$}   \\ \hline
 & This work & ref. \cite{pdg} & This work & ref. \cite{pdg} & This work &
ref. \cite{pdg}   \\ \hline
$\alpha _\pi (t)$ & ${\bf 135}$ & $135$ & ${\bf 1677\pm 8}$ & $1677\pm 8$ &
$2237\pm 26$ &       \\ \hline
$\alpha _K(t)$ & ${\bf 493.7}$ & $493.7$ & ${\bf 1773\pm 8}$ & $1773\pm 8$ &
$2333\pm 27$ &       \\ \hline
$\alpha _{\eta _s}$ & $698\pm 14$ &   & $1869\pm 38$ & $1854\pm 20$ & $2429\pm
54$ &       \\ \hline
$\alpha _D(t)$ & ${\bf 1864.1\pm 1.0}$ & $1864.1\pm 1.0$ & $2692\pm 19$ &   & 
$3228\pm 22$ &       \\ \hline
$\alpha _{D_s}(t)$ & $1971\pm 19$ & $1969.0\pm 1.4$ & $2786\pm 26$ &   &
$3323\pm 32$ &       \\ \hline
$\alpha _{\eta _c}(t)$ & ${\bf 2979.8\pm 2.1}$ & $2979.8\pm 2.1$ & $3692\pm 
23$ &   & $4217\pm 25$ &       \\ \hline
$\alpha _B(t)$ & ${\bf 5279.8\pm 1.6}$ & $5279.8\pm 1.6$ & $5830\pm 89$ &   & 
$6286\pm 93$ &       \\ \hline
$\alpha _{B_s}(t)$ & ${\bf 5369.6\pm 2.4}$ & $5369.6\pm 2.4$ & $5920\pm 89$ &
 & $6376\pm 93$ &       \\ \hline
$\alpha _{B_c}(t)$ & $6283\pm 79$ &   & $6826\pm 79$ &   & $7287\pm 80$ &   
 \\ \hline
$\alpha _{\eta _b}(t)$ & ${\bf 9424\pm 3.6}$&   & $9914\pm 148$ &   & $10353\pm
150$ &       \\ \hline
\end{tabular}
\end{center}
{\bf Table VI.} Comparison of the masses of the spin-0, spin-2 and spin-4
states given by ten pseudoscalar meson trajectories of the form (8) with 
data.\footnote{We take the error estimate on the ${\eta _b}$ mass as 10\% of 
the calculated splitting, in agreement with Fig. 2 of the second paper of ref.
\cite{NRQCD}.} All masses are in MeV.
 \\

\subsection{Axial-vector meson trajectories}
Finally,  we calculate the parameters of the axial-vector trajectories. 
These have the same parity and spin as the parity partners of the 
pseudoscalar trajectories -- the pseudovector trajectories\footnote{We use the 
usual Regge terminology for parity partners, which is not the same as normally 
used in lattice QCD/chiral condensates.}. The available data are 
insufficient to fix both thresholds and intercepts of axial-vector  
trajectories.  We assume, therefore, in  analogy with the tensor vs.  vector  
meson trajectories, that the thresholds of the parity partners coincide, and 
further, that the thresholds do not depend on charge conjugation in accordance
with the $C$-invariance of QCD. Under this assumption, 
we use the calculated thresholds of pseudoscalar trajectories and  fit only 
the intercepts of the  axial-vector  trajectories using the masses of a few 
lowest lying states and/or the additivity requirements, Eqs. (26),(27).

In particular, we use masses of $a_1$,  $D_{s1}$, $\chi _{c1}(1P) $ and $\chi 
_{b1}(1P) $. Note, that to find the parameters of trajectories of mesons 
containing strangeness, we use in the absence of data for $K_1$ the mass of 
$D_s$ rather than $f_1^{'}$. This is because the $f_1^{'}$ is not a pure 
$s\bar{s}$ state, and therefore the value of the intercept extracted from the 
physical state could deviate significantly from the value corresponding to a 
pure $s\bar{s}$ trajectory. (An additional complication arises, in principle,  
for axial mesons since the physical $q\bar{q'}$ states are mixtures of 
axial-vector and pseudovector components, because charge conjugation is not 
well-defined for other than $q\bar{q}$ states. Note that for all states other 
than axial-vector mesons, the corresponding state with opposite $C$ is exotic,
so that the problem is avoided. Of the inputs we use, this is relevant only 
to $D_{s1}$. We assume that the masses of $D_{s1A}$ and $D_{s1B}$ are close, 
because the heavy quark content implies that the hyperfine mass splitting is 
small.)

As before,  $\gamma $ is taken as universal, and the value found from the 
$\rho$ trajectory is used. The parameters we obtain in this 
way are in  Table VII.
\begin{center}
\hspace*{-1cm}
{\footnotesize
\begin{tabular}{|c|c|c|c|c|} \hline
  & $a_1 $ & $K_1 $ & ${f_1^{'}} $ & 
\\ \hline 
$\alpha (0)$ & $-0.03\pm 0.07$ & $-0.22 \pm 0.08$ & $-0.42 \pm 0.15$ & 
 \\ \hline
       & $D_1 $ & $D_{s1}$ & $\chi _{c1}(1P) $ & 
\\ \hline
$\alpha (0)$ & $-1.83\pm 0.05$ & $-2.03 \pm 0.06$ & $-3.63 \pm 0.07$ & 
 \\ \hline
 & $B_1^\ast $ & $B_{s1}$ & $B_{c1}$ & $\chi _{b1}(1P)$     \\ \hline
 $\alpha (0)$ 
 &$-7.87\pm 0.27$ & $-8.06 \pm 0.28$ & $-9.67 \pm 0.27$ & $-15.70 \pm 0.54$  \\  
\hline
\end{tabular}
}
\vskip0.2in
{\bf Table VII.} Parameters of axial-vector meson trajectories of the form (8).
 \\
\end{center}

Our results for masses of the states lying on the resulting  trajectories 
are compared with data from \cite{pdg} in Table VIII. The masses of the states 
for which there are no data available should, as before, be considered as our 
predictions.

\begin{center}
\begin{tabular}{|c|cc|cc|cc|} \hline
 & \multicolumn{2}{c|}{$J=1$} & \multicolumn{2}{c|}{$J=3$} &
 \multicolumn{2}{c|}{$J=5$}   \\ \hline
 & This work & ref. \cite{pdg} & This work & ref. \cite{pdg} & This work &
ref. \cite{pdg}   \\ \hline
$\alpha _{a_1}(t)$ & ${\bf 1230\pm 40}$ & $1230\pm 40$ & $2000\pm 31$ &   &
$2427\pm 32$ &       \\ \hline
$\alpha _{K_{1A}}(t)$ & $1368\pm 46$ &  & $2109\pm 33$ &   & $2535\pm 33$ & 
     \\ \hline
$\alpha _{f_1}(t)$ & $1501\pm 78$ & $1518\pm 5$ & $2218\pm 62$ &   & $2643\pm
67$ &       \\ \hline
$\alpha _{D_1}(t)$ & $2418\pm 26$ & $2422.2\pm 1.8$ & $3042\pm 25$ &   & $3473
\pm 24$ &       \\ \hline
$\alpha _{D_{s1}}(t)$ & ${\bf 2535.4\pm 0.3}$ & $2535.4\pm 0.3$ & $3150\pm 34$
 &    & $3580\pm 37$ &       \\ \hline
$\alpha _{\chi _{c1}(1P)}(t)$ & ${\bf 3510.5\pm 0.1}$ & $3510.5\pm 0.1$ & 
$4080\pm 30$ &    & $4513\pm 28$ &       \\ \hline
$\alpha _{B_1}(t)$ & $5692\pm 104$ &   & $6171\pm 102$ &   & $6570\pm 104$ & 
     \\ \hline
$\alpha _{B_{s1}}(t)$ & $5796\pm 105$ &   & $6273\pm 104$ &   & $6671\pm 105$
 &       \\ \hline
$\alpha _{B_{c1}}(t)$ & $6740\pm 95$ &   & $7215\pm 92$ &   & $7620\pm 90$ &
     \\ \hline
$\alpha _{\chi _{b1}(1P)}(t)$ & ${\bf 9891.9\pm 0.7}$ & $9891.9\pm 0.7$ & 
$10333\pm 173$ &   & $10727\pm 172$ &        \\ \hline
\end{tabular}
\end{center}
{\bf Table VIII.} Comparison of the masses of the spin-1, spin-3 and spin-5
states given by ten axial-vector meson trajectories of the form (8) with 
data. All masses are in MeV.
 \\

\subsection{Discussion of the results}
Here we comment on the results presented in this section:
\subsubsection{Intercepts}
\indent {\it Light-quark $(n\bar{n})$ trajectories:}\\
\noindent
i) The intercept of the linear trajectory that passes through $\rho$ and 
$\rho_3$, $\alpha(t) = 0.48 + 0.88 t$, should be reduced for the $a_2$ if the
$a_2$ were on a parallel  
linear trajectory: $0.47 + 0.88 t$. Even lower value for the intercept (and a 
different value of the slope), $0.30+0.98\;\!t$, is needed if $a_2$ and $a_4$ 
masses were to lie on the same linear trajectory (the latter being 
$1.944$ MeV, in accord with VES's recent measurements \cite{VES1}). (Note the 
discrepancy between the two sets of the $a_2$ trajectory parameters.) In our 
case, due to nonlinearity the intercept increases to 
$0.55$ for $\rho$ and $0.60 \pm 0.03$ for $a_2$. The data on $pp$, $\bar{p}p$ 
\cite{CKK} undoubtedly show that the tensor intercept is larger than the 
vector one (although the numerical errors of these data prevent us from using 
them as input). Therefore, data disfavor a linear trajectory.
\\
\noindent 
ii) The value $\alpha _{a_2}(0)=0.60 \pm 0.03$ that we obtain here is in 
agreement 
with the recent reanalysis by Cudell, Kang and Kim \cite{CKK} of the $pp$ and 
$\bar{p}p$ scattering data in the simple pole exchange model \cite{DL}
$\alpha _\rho (0)=0.50\pm 0.07,$ $\alpha _{a_2}(0)=0.66\pm 0.08.$
In this analysis the pairs of trajectories, $\rho $ and $\omega ,$ and $a_2$ 
and $f_2,$ are taken to be degenerate.  However, since $M(\omega )>M(\rho )$ 
and $M(f_2)>M(a_2),$ the intercepts of these trajectories should actually 
satisfy $\alpha _\omega (0)<\alpha _\rho (0)$ and $\alpha _{a_2}(0)<\alpha _{
f_2}(0).$ Hence, the value of $\alpha _\rho (0)$ should be larger the average 
of $\alpha _\rho (0)$ and $\alpha _\omega (0),$ i.e., 0.50. Similarly, the 
value of $\alpha _{a_2}(0)$ should be less than 0.66.

Our values for the intercepts agree well with their values  within the errors, 
moreover, they satisfy the above described inequalities. 
\\
\noindent
iii) Our value $\alpha _{a_1}(0)=-0.03 \pm 0.07$ is consistent with $\alpha 
_{a_1}(0)=-0.1\pm 0.2$ found in the analysis of the reaction 
$\pi ^{-}p\rightarrow \pi ^{-}\pi ^{+}n$ \cite{Alex}.

{\it Trajectories containing strangeness:}\\
\noindent
The values of $K^{\ast}$ and $K_2^{\ast}$ intercepts are in excellent 
agreement with the analysis of hypercharge exchange processes 
$\pi ^{+}p\rightarrow K^{+}\Sigma ^{+}$ and 
$K^{-}p\rightarrow \pi ^{-}\Sigma ^{+}$ \cite{VKKT}.

{\it Trajectories containing  a heavy quark:}\\
\noindent
i) The calculated intercepts of the heavy quark trajectories are 
negative, and the absolute value of an intercept increases with the increasing 
mass of the heavy quark. This feature is confirmed by means of the operator
expansion of QCD and dispersion relations for heavy quark four-current 
correlation functions by Oganesyan and Khodzhamiryan \cite{OK}. They also find
the estimates $\alpha _{J/\psi }(0)\sim -(2-3),$ $\alpha _\Upsilon (0)<-10.$ 
Our corresponding results, $\alpha _{J/\psi }(0)=-2.60 \pm 0.1$ and   $\alpha 
_\Upsilon (0) = -14.81 \pm 0.35$, agree very well with these estimates 
provided by QCD.\\
\noindent  
ii) Our predictions for the intercepts of the charmed meson trajectories may 
be confronted with the existing data. For the reaction
\begin{eqnarray*}
\pi^{-}A\rightarrow D(D^\ast )X
\end{eqnarray*}
the differential cross section, as given in the triple-Regge limit 
\cite{Muller}, is
\begin{eqnarray*}
\frac{d^2\sigma }{dx_Fdp_\perp ^2}=A(1-|x_F|)^n e^{-bp_\perp ^2},\;\;\;A,b=
{\rm const},
\end{eqnarray*}
where $x_F$ and $p_\perp $ are the Feynman-$x$ and transverse momentum 
variables, respectively, and $n=1-2\alpha (0),$ with $\alpha (0)$ being the 
intercept of the trajectory exchanged in the $t$-channel of the reaction. 
The $D$ meson production proceeds via the exchange of both $D^\ast $ and 
$D_2^\ast $ trajectories. Hence, on the basis of our results for the 
corresponding intercepts, one may expect $n\simeq 3.2\pm 0.2$ for the 
$D$-production. In $\pi ^{-}$H interactions at 360 GeV the NA27 experiment by 
the LEBC-EHS collaboration \cite{NA27} finds $n=3.80\pm 0.63.$ For 350 GeV 
$\pi ^{-}$ beam on emulsion the WA75 collaboration \cite{WA75} finds $n=3.5
\pm 0.5.$ In $\pi ^{-}$Cu interactions at 230 GeV the ACCMOR collaboration 
\cite{ACCMOR} finds $n=3.23\pm 0.29$ for leading charm production (and 
$n=4.34\pm 0.35$ for nonleading, and $n=3.74\pm 0.23$ for the combined data). 
For 340 GeV $\pi ^{-}$ beam on Si, Cu and W targets the WA82 collaboration 
\cite{WA82} finds $n=2.9\pm 0.3.$ Finally, for 250 GeV $\pi ^{-}$ beam on Be, 
Al, Cu and W targets the E769 collaboration \cite{E769} finds $n=3.2\pm 0.5$ 
on Al target, and $n=3.7\pm 0.4$ for leading $D$ production on all targets 
(and $n=4.0\pm 0.4$ for nonleading, and $n=3.9\pm 0.3$ for the combined data). 

Since the $D^\ast $ meson production can proceed via the exchanges of all 
four, $D^\ast,$ $D_2^\ast ,$ $D$ and $D_1$, trajectories calculated here, 
definite predictions for 
$n$ are more complicated than in the previous case. Of course,  
quantitatively $n$ should be larger in this case, because the intercepts of 
$D$ and $D_1$ trajectories are larger in magnitude than those of $D^\ast $ and 
$D_2^\ast $ trajectories. In fact, in the experiment mentioned above the E769 
collaboration \cite{E769-2} finds $n=3.5\pm 0.3.$ Earlier measurements in the 
NA27 experiment \cite{NA27} produced $n=4.3^{+1.8}_{-1.5}.$

The value $\alpha _{D^\ast }(0)=-1.02\pm 0.05\approx -1$ obtained in our 
analysis also supports the use of the $(1-|x_F|)^3$ production model which was 
claimed to describe the available data reasonably well \cite{E515}. 

\subsubsection{Spectroscopy}
i) {\it Masses of $s\bar{s}$ states.}
Our predictions for the masses of pure 
$s\bar{s}$ states cannot be directly compared to  the values for physical
states quoted by \cite{pdg}, because the physical states emerge upon mixing 
with the nonstrange isoscalar states of the corresponding meson multiplets. 
Comparison should be therefore made with the pure $s\bar{s}$ states 
calculated from both the established physical $n\bar{n}$ and $s\bar{s}$ 
states. Pure $s\bar{s}$ states were calculated in \cite{BGP} with the 
help of Schwinger's quartic mass formula, for both linear and quadratic 
masses, in essential agreement in both cases and consistent with the 
quark model motivated linear mass relation   
$M(n\bar{n})+M(s\bar{s})=2M(s\bar{n}).$ 
Here we quote the masses of the pure $s\bar{s}$ states found in \cite{BGP} 
(specifically, from Schwinger's formula for quadratic masses)
for vector $J=1,3$, tensor $J=2$ and pseudoscalar $J=2$ trajectory multiplets, 
and calculate the pure $s\bar{s}$ mass for the axial-vector $J=1$ multiplet as
well (all masses are given in MeV). We find  that these values and the 
corresponding values in Tables II,IV,VI,VIII are in essential agreement:

$\bullet $ Vector: $J=1$ $M(s\bar{s})=1014.5\pm 0.4,$ $J=3$  
$M(s\bar{s})=1862.3\pm 9.0$

$\bullet $ Tensor: $J=2$ $M(s\bar{s})=1539.4\pm 5.5$

$\bullet $  Pseudoscalar: $J=2$ $M(s\bar{s})=1869.1\pm 25.3$

$\bullet $ Axial-vector: $J=1$ $M(s\bar{s})=1487.9\pm 30.0$

Interestingly enough, if one applies the quark model motivated relation
mentioned above to calculate the mass of the axial-vector $s\bar{n}$ state 
using the predicted pure $s\bar{s}$ mass (this work)  and  the physical $a_1$ 
(as pure $n\bar{n})$ 
mass (ref. \cite{pdg}) as inputs, one obtains $M(s\bar{n})=1359\pm 25$ MeV, in 
good 
agreement with the corresponding value in Table IV.

ii) {\it A new $c\bar{u}$ meson, $D(2637),$} with mass $2637\pm 2\pm 6$ MeV, 
was recently reported by the DELPHI collaboration \cite{DELPHI}. Page 
\cite{Page} has demonstrated from heavy quark symmetry that the width of 
$D(2637)$ claimed by DELPHI is inconsistent with any bound state with one 
charm quark predicted in the $D(2637)$ mass region, except possibly $D_3^
\ast ,$ $D_2$ or $2S$ $D,$ among the three $D_2$ being the best candidate. 
Interestingly enough, the mass of $D_2$ predicted in Table VI, $2692\pm 19$ 
MeV, is consistent with the value of the $D(2637)$ mass measured by DELPHI. 

iii) {\it  Our predictions for the mass of the $B_c,$ $B_c^\ast $ states} 
are in a good agreement with the predictions of different models compiled by 
Kiselev {\it et al.} \cite{KLT} which all, including their own prediction, lie
in the ranges $6281\pm 38$ and  $6343\pm 26$ MeV, respectively. 

The $B_c$ mass that we obtained, $6283\pm 79$ MeV, is also consistent with the 
value $6.40\pm 0.39\pm 0.13$ reported by the CDF collaboration \cite{CDF}.

iv) {\it The value of the $B_1$ meson mass} that we obtained, $5692\pm 104$ 
MeV, is in good agreement with that recently reported from experiment 
\cite{Muheim}: $5675\pm 12\pm 4$ MeV.

v) {\it States near the trajectory termination point.}
As we argued in Sections 2 and 3, any nonlinear trajectory of the form $(5)$, 
including the square-root, can be expected to fit well the lowest lying 
states, and discrepancies are likely to arise only for a few states near the 
trajectory threshold. There are not enough data to confirm or deny this  
statement, with the possible exception of the $a_2$ trajectory. Our prediction
for the mass of $a_6$, which may be the last state on the true $a_2$ 
trajectory, and is certainly next-to-last on our corresponding square-root 
trajectory (see discusion in Section 4.2), is 10\% lower than the observed 
value. This is in contrast to the $a_4$ mass which is lower only by about 1\%.
The same is true for the $K_4^\ast $ mass. (Both $a_4$ and $K_4$ lie about 550
MeV below the corresponding thresholds, whereas the observed mass of the $a_6$
is right at our threshold, in contrast to the calculated value which is 
$\sim 200$ MeV below threshold.)

Even though the agreement within the errors is acceptable, in our opinion, 
this raises the  possibility that near the termination point, the true hadron 
trajectory grows slower than the square-root trajectory. If this possibility 
is realized, the $K_6^\ast $ is likely to lie about 10\% above the value 
we predict, and it can be expected to be broad, similar to the observed $a_6$.

It should be mentioned that there are other trajectories for which the second 
state on the trajectory is known, e.g. some of the vector or pseudoscalar 
trajectories. There are two factors which prevent observing the possible 
general tendency of the square-root form to overshoot the true trajectory. 
First, these states are further away from the threshold than those discussed 
above, in particular, they are about 770 MeV below threshold for the 
vector trajectories, and about 1200 MeV for the pseudoscalars. This means that
the effect can be expected to be even less than 1\%. Second, in both of these 
cases, we used some of the second states on the trajectories as inputs, thus 
achieving better agreement with more of the higher mass states on a trajectory.
 
vi) {\it Additivity constraints and mass relations.}
Our results for masses are in excellent agreement  with available data. This 
is not suprising, however, because the additivity constraints (26) and (27) 
are known to lead to high-accuracy higher-power mass relations \cite{BGH}. 
Therefore, the numerical spectroscopy results reported here should be viewed 
solely as a confirmation that the particular square-root form of trajectory 
does not introduce any systematic biases.

\subsubsection{A consequence of trajectory thresholds}
If the hadronic Regge trajectories indeed terminate in  $\ell $, an intriguing 
possibility that might simplify  identification of states arises. Even though 
the maximum $\ell $ is very sensitive to the specifics of the potential 
(in particular, its long distance behaviour) and cannot be determined 
at all by fitting the bound states, the  thresholds turn out to be insensitive 
to how the potential approaches its asymptotic value and very well  
determined  by the bound state spectra. Recall our numerical simulation of 
the model in Section 2: The bound state spectra were well approximated by both 
extreme cases, the square-root and the logarithmic trajectory. These two forms 
differ drastically in the maximum $\ell $ (finite vs. infinite), nevertheless, 
they both predict $T$ within a few percent of the model's true value. 
We therefore conjecture that one can predict ``spectroscopy windows'', i.e. 
range of masses for each particular flavor. Even though we  cannot obtain 
stringent error estimates on these predictions, it is conceivable that the 
deviations from the true values are small. 

The most intriguing consequence of this conjecture is, in our opinion, the 
following: According to our calculations, there are no light quarkonia states 
beyond about $3.2$ GeV. (The highest threshold for light quarkonia is 
$\sqrt{T_{\eta_s}}= 3.10 \pm 0.11$ GeV.) Even though the charmed states start 
around 3 GeV, their mixing with glueballs is small. Therefore, if our analysis
is right, any state above 3 GeV (to, say, 5 GeV) that does not fit into the 
charmed spectra can be expected to be predominantly a glueball or an exotic. 

\section{Summary and Conclusions}
Previously \cite{BBG} we argued that in QCD, the real part of hadronic Regge 
trajectories should acquire curvature and terminate as a consequence of the 
flux tube breaking due to pair creation. In this paper we addressed the 
issue of what is the specific form of the trajectories. We started with the 
simple potential model of two heavy quarks in a potential which is screened at 
large distances, and we have shown that the parent trajectory formed by the 
bound states of the system can be equally well approximated by the square-root 
and the logarthmic forms. This is not suprising, once the comparison between 
the potentials which lead to these trajectories, respectively, and the 
screened potential of the heavy quark model is made. Nevertheless, even though
the first two are indistinguishable, the square-root form is closer to the 
real trajectory because its real part terminates (as does the real part of the
heavy quark model), while the logarithmic trajectory grows without bound. This
growth is specific to the logarithmic form; any other trajectory of our 
nonlinear form (5) with $\nu \neq 0$ has a termination point. This means that 
any $\nu \neq 0$ would approximate the true trajectory both quantitatively and
qualitatively. Different forms can be expected to lead to subtle differences 
for higher excited states, particularly for those near the threshold. 
Until a distinction can be made, we choose the 
square-root trajectory to study real hadronic spectra. A nice feature of this 
trajectory is that the additivity of inverse slopes reduces to a simple 
expression which has natural units of mass, and that the parameter $\gamma $ 
is given simply in  units of  inverse mass. It is possible that any of these 
forms, including the square-root, grow faster near the threshold than the
true trajectory. It is also plausible, although not conclusive, that an 
indication of this tendency has been seen in the $a_2$ trajectory. This is 
also supported by the rate of growth of the lattice potential as compared to 
the potential which leads to the square-root trajectory, although the lattice 
potential should not be taken too seriously at large distances. 

With the form  of trajectory chosen, the parameters need to be fitted. We have 
typically used masses of few lowest lying states, and/or the intercepts if 
known and reliable, such as the intercept of the $\rho $ trajectory. We also 
utilized the additivity requirements. Our calculation is in an excellent 
agreement with various data, both spectroscopic and scattering, and it is 
self-consistent (see the discusion of pseudoscalar $b$-mesons).

We conclude that we have provided both strong phenomenological arguments and  
theoretical considerations which indicate that the hadronic Regge trajectories 
are essentially nonlinear, and that they can be well approximated by the 
square-root form (at least for the lowest lying states). 
This observation has a profound effect on our theoretical understanding of 
hadronic spectra, namely, it implies that the linear confinement is not 
sufficient and most important factor  determining the position of poles, not 
even for states as low as the second or third on a trajectory. Perhaps an even
more significant consequence of the nonlinearity of trajectories, and, in 
particular, the existence of the trajectory thresholds, is that it may 
simplify identification of states in experiements. For example, once the 
existence of thresholds is established, it would be easier to identify exotic 
states as those that do not fit in the heavy-quarkonia spectra.  
 
\section*{Acknowledgments}
We would like to thank P.R. Page for reading the manuscript, and comments and 
suggestions. One of us (L.B.) wishes to thank L.P. Horwitz for discussions, 
and K.V. Vasavada for correspondence, during the preparation of this work. 

This research was supported by the Department of Energy under contract
W-7405-ENG-36.

\newpage
\begin{appendix}
\catcode`\@=11 \@addtoreset{equation}{section}

\renewcommand{\theequation}{A.\arabic{equation}}  
\section{The allowed values of $\nu$ }
It can be shown that for $n<\nu <n+1$ (for integer $\nu ,$ $\alpha (t)
={\rm Re}\;\alpha (t))$, the trajectory of the form (5) 
satisfies the following dispertion relation with $n+1$ subtractions:
\begin{equation}
\alpha (t)=\alpha (0)+\alpha'(0)\;\!t+\ldots 
+\frac{\alpha^{(n)}(0)t^n}{n!}+\frac{t^{n+1}}{\pi }\int _T^\infty 
dt'\;\!\frac{{\rm Im}\;\!\alpha (t')}{t'^{n+1}(t'-t)}.
\end{equation}
Since for the trajectory (5) ${\rm Im}\;\!\alpha (t)=\sin (\pi \nu ) 
(t-T)^\nu ,$ the requirement of the positivity of the trajectory 
imaginary part (this requirement follows from unitarity \cite{LASZLO}
leads to $2k<\nu <2k+1,$ where $k$ is integer. 

For the trajectory parametrization 
\begin{equation}
\alpha (t)=-\gamma (-t)^\nu [\log (-t)]^\beta 
\end{equation} 
(which is the $t>>T$ limit of (5), up to a power of logarithm) 
it was shown in \cite{Tru} that $0\leq \nu \leq 1$ and $0<\beta \leq 2.$

The range of $\nu $ can be  further restricted.
The class of dual models called dual amplitudes with Mandelstahm analycity 
(DAMA) \cite{DAMA} leaves open a corridor for possible asymptotic behavior of 
$\alpha (t),$ bounded from above by $\sqrt{|t|}$ and from below by 
$\log (|t|)$ \cite{LASZLO}. (DAMA has the Veneziano limit $\alpha (t)\sim t,$ 
but the transition to this limit occurs discontinuously \cite{LASZLO}.) Thus, 
in DAMA the range of $\nu $ is squeezed down to $0\leq \nu \leq 1/2.$

Hence we consider the value of $\nu$  
restricted to lie between $0$ and $1/2$ \cite{LASZLO}.  

\renewcommand{\theequation}{B.\arabic{equation}}  
\section{The dynamics of the generalized string model}
By varying the action of the generalized string with massive ends 
(here the dot stands for the derivative with respect to $\tau ,$ Lorentz 
invariant evolution parameter for the string),
\begin{eqnarray}
S_{gen}=\int \!\!\!\!\int d\tau ds  L(x,\dot{x},x')+\sum _{i=1,2}
L^{(m)}(\dot{x}_i),
\end{eqnarray}
one obtains the equations of motion of the generalized string,
\begin{eqnarray}
\frac{d}{d\tau }\frac{\partial L}{\partial \dot{x}}+\frac{d}{ds  }
\frac{\partial L}{\partial x'}=\frac{\partial L}{\partial x},
\end{eqnarray}
and the boundary conditions which represent the equations of motion of the
massive ends:
\begin{eqnarray}
\frac{d}{d\tau }\frac{\partial L^{(m)}}{\partial \dot{x}_i}=\frac{
\partial L}{\partial x'},\;\;\;x=x_i.
\end{eqnarray}

In the gauge $\tau =t$ discussed above, the equations of motion of the 
generalized string reduce to
\begin{eqnarray}
\frac{d}{dt}\frac{\partial L}{\partial \dot{x}}+\frac{d}{ds  }\frac{
\partial L}{\partial x'}=\frac{\partial L}{\partial x},\;\;\;x\equiv {\bf x},
\end{eqnarray}
and the boundary conditions are
\begin{eqnarray}
m_1\;\!\frac{d}{dt}\frac{\dot{x}_1}{\sqrt{1-\dot{x}_1^2}} & = & \frac{
\partial L}{\partial x'},\;\;\;s  =0, \nonumber \\
m_2\;\!\frac{d}{dt}\frac{\dot{x}_2}{\sqrt{1-\dot{x}_2^2}} & = & \frac{
\partial L}{\partial x'},\;\;\;s  =\pi . 
\end{eqnarray}

Let us show that, similarly to the standard case of the string with constant
tension, there are solutions to the equations of motion of the generalized
string (with the Lagrangian given in (14)) in the form of a rigid rod connecting 
the massive ends and rotating 
with frequency $\omega $ about its center of mass:
\begin{eqnarray}
x(t,s  )=\rho (s  )\Big( \cos (\omega t),\;\sin (\omega t),\;0\Big) .
\end{eqnarray}
Indeed, $\sigma =\sigma (|x|)=\sigma (\rho ),$ since $x^2=\rho ^2;$ therefore 
$d\rho/dx=x/\rho =(\cos (\omega t),\;\sin (\omega t),\;0),$ and $d\sigma /dx=d
\sigma /d\rho $ $d\rho /dx=d\sigma /d\rho $ $(\cos (\omega t),\;\sin (\omega 
t),\;0).$ Hence 
\begin{eqnarray}
\frac{\partial L}{\partial x}=\frac{\partial L}{\partial \sigma }\;\!
\frac{d\sigma}{dx}=-\frac{d\sigma }{d\rho }\;\!\rho' \sqrt{1-\omega ^2
\rho ^2}\Big( \cos (\omega t),\;\sin (\omega t),\;0\Big) .
\end{eqnarray}
Since also
\begin{eqnarray}
\frac{d}{dt}\;\!\frac{\partial L}{\partial \dot{x}}=-\frac{\sigma \omega ^2
\rho \rho'}{\sqrt{1-\omega ^2\rho ^2}}\;\!\Big( \cos (\omega t),\;\sin 
(\omega t),\;0\Big) ,
\end{eqnarray}
\begin{eqnarray}
\frac{d}{ds  }\;\!\frac{\partial L}{\partial x'}=\left( -\frac{d\sigma }{
d\rho }\;\!\rho' \sqrt{1-\omega ^2\rho ^2}+\frac{\sigma \omega ^2\rho \rho'}{
\sqrt{1-\omega ^2\rho ^2}}\right) \Big( \cos (\omega t),\;\sin (\omega t),
\;0\Big) ,
\end{eqnarray}
(the last relation is obtained via $d\sigma /ds  =d\sigma /d\rho $ 
$\rho'),$ it follows that the equations of motion (B.4) are satisfied. 

One can show that for the rotation (B.6), the energy of the 
generalized string is given by
\begin{eqnarray}
H=\int ds  \sqrt{p^2+\sigma ^2x'^2}=\int ds  \;\!\frac{\sigma 
\rho'}{\sqrt{1-\omega ^2\rho ^2}}=\int \frac{d\rho \;\!\sigma (\rho )}{
\sqrt{1-\omega ^2\rho ^2}}.
\end{eqnarray}
Similarly, the orbital momentum of the generalized string is 
\begin{eqnarray}
J=J_z=\int ds  \left( xp_y-yp_x\right) =\int ds  \;\!\frac{\sigma 
\omega \rho ^2\rho'}{\sqrt{1-\omega ^2\rho ^2}}=\int \frac{d\rho \;\!
\sigma (\rho )\omega \rho ^2}{\sqrt{1-\omega ^2\rho ^2}}.
\end{eqnarray}
Interestingly enough, in his book \cite{Per} Perkins also presents the above 
relations for the energy and orbital momentum of the generalized string. He
does not however derive these relations from the first principles Lagrangian,
as in Eq. (14). 

By adding the contribution of the massive ends, one finally has the 
expressions for the total energy and orbital momentum of the generalized 
string with massive ends:
\begin{eqnarray}
E=\int _0^{r_1}\frac{d\rho \;\!\sigma (\rho )}{\sqrt{1-\omega ^2\rho ^2}}+
\int _0^{r_2}\frac{d\rho \;\!\sigma (\rho )}{\sqrt{1-\omega ^2\rho ^2}}+
\frac{m_1}{\sqrt{1-\omega ^2r_1^2}}+\frac{m_2}{\sqrt{1-\omega ^2r_2^2}},
\end{eqnarray}
\begin{eqnarray}
J=\int _0^{r_1}\frac{d\rho \;\!\sigma (\rho )\omega \rho ^2}{\sqrt{1-\omega ^2
\rho ^2}}+\int _0^{r_2}\frac{d\rho \;\!\sigma (\rho )\omega \rho ^2}{\sqrt{1-
\omega ^2\rho ^2}}+\frac{m_1\omega r_1^2}{\sqrt{1-\omega ^2r_1^2}}+\frac{
m_2\omega r_2^2}{\sqrt{1-\omega ^2r_2^2}}.
\end{eqnarray}
Note that the boundary conditions (B.5) define the separations of the massive 
ends from the center of mass through the following nonlinear equations:
\begin{eqnarray}
\frac{m_i\omega ^2r_i}{\sqrt{1-\omega ^2r_i^2}}=\sigma (r_i)\sqrt{1-\omega ^2
r_i^2},\;\;\;i=1,2.
\end{eqnarray}
\end{appendix}

\newpage

\end{document}